# Topological Laser in Anomalous Quadrupole Topological Phases

*Guangtai Lu\*, Yasutomo Ota and Satoshi Iwamoto*

Email: lugt@iis.u-tokyo.ac.jp

G. Lu, S. Iwamoto
Research Center for Advanced Science and Technology, The University of Tokyo, 4-6-1 Komaba, Meguro-ku, Tokyo 153-8904, Japan

Y. Ota
Department of Applied Physics and Physico-Informatics, Faculty of Science and Technology, Keio University, 3-14-1 Hiyoshi, Kohoku-ku, Yokohama-shi, Kanagawa 223-8522, Japan

**Abstract**
Topological photonics shows considerable promise in revolutionizing photonic devices through the use of topological phases, leading to innovations like topological lasers that enhance light control. One of recent breakthroughs is reducing the size of these systems by utilizing lower-dimensional boundary states, notably via higher-order topological phases. This paper presents the first experimental demonstration of topological laser in anomalous quadrupole topological phase, an instance of higher-order phases. To facilitate this, a topological nanocavity with quality factor near 6,000 is engineered through a twisting operation. The topological nature of our system is validated by calculation of nested Wannier center and the emergency condition of corner states. Our experimental observations reveal the manifestation of corner states and the achievement of single-mode pulsed laser, driven by optical gain from multiple quantum wells at telecommunication wavelengths and at a temperature of 4 K. A lasing threshold of 23 µW and a cold quality factor of 1,500 are deduce through rate equation. Our work gives a new potential in the application of topological principles to advance nanophotonic technologies.

## 1. Introduction

Topological photonics is proving to be highly effective in advancing photonic systems.[1–3] Due to the variety of topological phases, plenty of novel photonic devices utilizing different topological states have been demonstrated, including unidirectional photonic waveguides with photonic quantum Hall phase,[4–6] robust topological waveguides with photonic quantum spin Hall phase[7–10] and photonic quantum valley Hall phase,[11–13] and topological nanocavities with higher-order topological phase.[14–31] The integration of topologically protected states with gain materials has sparked considerable interest, leading to the development of topological lasers.[20–22,32–44] These lasers offer novel ways to manipulate light, such as unidirectional emitting[34,37], and contributes to the evolution of laser technology. Early investigations predominantly focused on harnessing the resilience of one-dimensional edge states within two-dimensional (2D) systems.[34–40] However, the inherent nature of these edge states resulted in lasing systems with substantial footprints. Some efforts on the zero-dimensional (0D) topological states in one-dimensional (1D) photonic crystals helps to deduce the mode volume of topological lasers.[44] A groundbreaking advance in higher-order topology[14–17] has introduced a new method for achieving topological lasers with minimized mode volumes, employing lower-dimensional boundary states. Such systems exhibit lower-dimensional boundary states, notably, the emergence of 0D corner states in 2D configurations mainly characterized by 2D Zak phase or quantized quadrupole moment. The systems characterized by 2D Zak phase[18,20–22] have gained traction due to its ease of implementation in compact, magnet-free optical platforms, making it a viable choice for lasing systems. In contrast, the untapped potential of semiconductor photonic crystals characterized by quantized quadrupole moments has hindered its incorporation into lasing systems.

Systems characterized by quantized quadrupole moments, often referred to as being in quadrupole topological phases, demonstrate a phenomenon where bulk quadrupole moments induce confined corner charges in 2D finite materials. While the original definition of quadrupole moments pertains to electronic systems, a mathematical generalization into other neutral systems described by wave equations has paved the way for photonic quadrupole topological insulators (QTIs). The early proposal of photonic QTIs requires negative coupling coefficient to establish π-flux lattices.[14,15] Following this idea, photonic QTIs have been observed in several platforms, including microwave circuits[23], coupled ring resonators[24], and coupled waveguides.[25]

Alternatively, magnetized systems have also been employed to generate quadrupole topological insulators.[26,27] However, these mechanisms face challenges in semiconductor photonic crystals (PhCs), primarily due to the difficulty in realizing complex coupling coefficients and weak magneto-optical effects in the optical regime. In our pursuit of a suitable platform for quadrupole topological lasers, we turn to the recently proposed anomalous quadrupole topological insulators (AQTIs).[28-30] Distinguished from conventional QTIs, AQTIs are protected by two orthogonal glide symmetries, eliminating the need for complex coupling strength or external magnetic fields. Despite the difference in fundamental mechanism, AQTIs exhibit quantized quadrupole moments and topologically protected corner modes similar to QTIs.

In this paper, we present the first demonstration and optical characterization of a laser based on a topological corner state in photonic AQTIs. We consider a corner composed of two semiconductor photonic crystal slabs, which are topologically non-trivial in the sense of quadrupole topology. Our study unveils the predictable formation of localized states at the corner as long as the two photonic crystals exhibiting different topological properties. The corner state has a high quality factor and serves as a photonic cavities. We experimentally fabricate the device using quantum wells with optical gain around telecommunication wavelength, and verify the existence of localized states. Through analysis, we establish a correlation between our numerical calculations and experimental measurements, affirming the observation of the corner state in photonic crystals. Furthermore, as optical gain increases, a clear lasing oscillation from the corner state is observed.

## 2. Results

### 2.1. Numerical Calculations

Figure 1a shows a design schematic of the quadrupole topological PhC investigated in this study. The unit cell with period $a = 850$ nm is composed of four air holes with radius $r = 144$ nm in InGaAsP-based 250 nm-thick slab (refractive index, $n = 3.4$). Four circular holes are originally located at ($\pm a/2$, $\pm a/2$), and are twisted by the same angle $\theta$ along the dashed lines.

The band structure using the three-dimensional finite element method (3D-FEM) for unit cell with $\theta = 6°$ is shown in Figure 1b. A large photonic band gap from 194 to 212

THz exists between fourth and fifth bands. To characterize the non-trivial quadrupole topology carried by this band gap, we calculate the so-called "nested Wannier bands". Following the discussion by W. Benalcazar, et al.[14], we first compute the Bloch functions of four bulk bands below the bandgap, and obtain the non-degenerate Wannier bands $v_x$ (Figure 1c) through Wilson loop method. There are four gapped Wannier bands corresponding to four bands below the band gap. We label them from 1 to 4 and notice that Wannier sectors I and II, representing combination of Wannier bands 1+3 and 2+4, lead to gapped composite Wannier bands. We then calculate the Wannier bands for Wannier sector I and II, which is also known as nested Wannier Bands $p_y^{v_x}$. The results, in Figure 1c, shows that the polarization of Wannier sector I and II are well quantized to ±1/2. Due to the $C_4$ symmetry in unit cell, the polarization of Wannier band in orthogonal direction shows the same value, which means $p_y^{v_x} = p_x^{v_y}$. And finally, the quadrupole moment is quantized as $q_{xy} = 2p_y^{v_x} p_x^{v_y} = 1/2$. One should notice that there is a phase transition at $\theta = 0°$ (Figure 1d), which indicating that PhCs with positive and negative twisting angle belong to different topological phase.

The topological phenomenon does not only express itself in Wannier space, but also in physical space, offering profound insights into the behavior of topological PhCs. Let us consider the structure composed of two PhCs, each characterized by independent twisting angles (Figure 2a). The principle of bulk-corner correspondence suggests that topological corner states emerge when the two PhCs reside in distinct topological phases. Figure 2b-h illustratively captures the evolution of eigenstates as a function of the twisting angle. By holding the twisting angle $\theta_1$ of the right-top PhC constant and progressively adjusting the twisting angle $\theta_2$ of the left-bottom PhC from negative to positive values, a critical observation is made: the corner state vanishes once $\theta_2$ surpasses 0 degrees. To quantify the localization of corner state, a localization strength function $\eta$ is introduced:

$$\eta = \frac{\int_{Corner} \varepsilon |\boldsymbol{E}|^2 dV}{\int \varepsilon |\boldsymbol{E}|^2 dV} \tag{1}$$

where $\varepsilon$ is the permittivity, $\boldsymbol{E}$ is the electric field, and $Corner$ represents an integral region near the center of device (yellow square in Figure 2c). $\eta$ represents the portion of energy localized near the corner region, thereby serving as an indicator for the localization strength of corner states. As shown in Figure 2b, the behavior of $\eta$ effectively maps the transition of states from being highly localized at the corner to merging into the bulk state. The state of interest decrease in localization strength $\eta$ as $\theta_2$ transitions from $-10°$ to $0°$ and remains significantly weak beyond $0°$, indicating the merger of corner states into bulk states. One may also notice that this process happens without gap closing[30], different from most of the topological phase transitions. Conversely, when we fix $\theta_2$ as $-6°$ and change $\theta_1$, the corner state only exist when $\theta_1$ is positive (see supporting information). This behavior supports the idea that PhCs with opposite signs of twisting angles exhibit differing topological phases, whereas those with matching signs of twisting angles align in the same topological phase. These findings not only corroborate with the predictions made by the nested Wilson loop calculations but also reinforce the concept that the corner states are inherently topological rather than mere conventional defects.

To maximize the bulk band gap, we study the structure composed of two PhCs with opposite twisting angle, which is $\theta_1 = -\theta_2 = \theta$. In such case, two PhCs share the same photonic band structure but are distinct in terms of band topology characterized by the polarization of Wannier sector. Figure 3a shows that such structure holds gapped edge states and in-gap corner state, as predicted by theory.[29,30] Figure 3b plots the quality factor of corner state as a function of twisting angle. The frequency and mode volume are summarized in Figure 3c. One notices that as twisting angle increases, corner states have the following three trends: (1) wavelength increases; (2) mode volume decreases; (3) the quality factor reaches its maximum value around 6,000 when $\theta = 4°$, and decrease thereafter. For small twisting angle, as shown in Figure 3d, the mode area is large, which leads to a strong light leakage through the simulation domain boundary. As twisting angle increases, the field profile of corner state shrinks and suppress the side leakage (Figure 3e-g). For twisting angle larger than 4º, the mode is so confined that large amount of components in momentum space are above the light cone, resulting in a strong out-of-plane leakage and thus a low quality factor.

## 2.2. Experimental Results

We fabricated the designed quadrupole topological photonic crystals in an air-suspended InGaAsP-based multiple quantum well layer using standard semiconductor nanofabrication process (see supporting information). The size of device is around 15 μm×15 μm, including 16×16 unit cells (8×8 unit cells with positive twisting angle and other unit cells with negative twisting angle) and 4 air windows for sacrificial layer etching. A scanning electron microscope image of the device with $\theta = 6°$ is shown in Figure 4a and 4b.

To characterize the fabricated devices, we performed the micro-photoluminescence (µPL) measurement at a low temperature of 4 K. The pump beam was generated by a diode laser oscillating at 830 nm (repetition rate 40 MHz, pulse width 6 ns), and focused on the surface of sample by an objective lens with a magnification of 50. PL signals were analyzed by a grating spectrometer and an InGaAs camera. Figure 3d shows the spectra measured with an average pump power of 8 µW for the devices with different twisting angle 4°, 6°, and 8°. We find a peak resonating in each spectrum. The center wavelength of the peaks show an increasing tendency as twisting angle increases (Figure 4c), which is coincident with our numerical calculation results. The sample with $\theta = 6°$ shows the narrowest linewidth and largest quality factor, which deviates slightly from our numerical calculations (maximum appears at $\theta = 4°$). This could possibly be due to the difference in boundary conditions. In experiments, modes may leakage through the sacrificial layer at the boundary because of the undercut shape (see Figure 4b). This may lead to the dominance of leakage at the boundary compared to out-of-plane radiation, causing the peak of quality factor to move to a larger twisting angle.

To ascertain that the observed peaks originate from corner resonant modes, further characterizations are performed on structures with a 6° twisting angle. Figure 4d shows the PL spectrum from a device composed of a single type of PhC. As expected from the bulk-corner correspondence, cavity states disappears, leaving the background emission from multiple quantum wells remains. This supports that the peaks we observed generates from the corner structure. We also measured the far field polarization and field distribution of the resonant mode. The far field polarization was measured by placing a linear polarizer in front of the spectrometer. Figure 4e shows the peak intensities resonant mode through the linear polarizer as its transmission axis is rotated. By fitting the data point, the polarization was found to be 57°, which matches with our numerical calculations (47°). To investigate the spatial extends of the resonant mode, we measured the pump position dependences of the PL spectrum, as shown in Figure 4f. It is clear that this resonant mode confined at the center of the device, which indicating that such mode localized at the corner. We further plot the field distribution of corner mode by numerical calculations as a comparison. After considering the pump spot, whose diameter is estimated to be 2 µm, the experiment data shows good agreement with numerical calculation. Supported by the above evidence, we conclude that the resonant mode we observed is the corner state.

Exploring the pump power dependence of the corner state in devices with a 6º twisting angle (Figure 5a), we observed a clear transition into lasing oscillation. As the pump power increases, the corner mode emission intensifies, dominating the spectrum. A sharp increase in output intensity beyond a threshold power provides clear evidence of lasing. The logarithmic light-in-light-out (LL) curve exhibits an S-like shape, and fitting through rate equations describe below.[45]

$$\frac{dp}{dt} = -kp + \beta\gamma(N - N_t)p + \beta\gamma N, \tag{2}$$

$$\frac{dN}{dt} = P - \gamma N - \beta\gamma(N - N_t)p, \tag{3}$$

where $p$ and $N$ are the cavity photon and carrier number. $k$ is the transparent cavity leakage rate of 800 GHz. $\gamma$ is the total spontaneous emission rate of 0.5 GHz. $P$ is the pump power. $N_t$ is the transparent carrier number, and is estimated to be 10,000 by multiplying active volume and transparent carrier density.[46] $\beta$ is the spontaneous emission coupling factor. Spontaneous emission rate is supposed to be linearly dependent on carrier number[47], and non-radiative recombination is not considered since the device was working at a low temperature. Owing to the long pulse width of 6 ns, we assume the time derivate to be zero to calculate the steady state. After fitting the LL curve, we deduce a $\beta$ factor of 0.15. This factor, orders of magnitude larger than those of conventional semiconductor lasers ( $\beta \sim 10^{-6}$ ), signifies tight optical confinement. We also find the threshold power is 23 μW, at which the net stimulated gain equals the loss. Figure 5b depicts the evolution of cavity linewidths and center wavelengths. We also find a significant blue shift of center wavelength due to carrier filling effect. A linewidth narrowing by nearly one order of magnitude is also observed, further confirming the lasing oscillation. Through the laser model used for the fitting, we estimate the transparent pump power to be 11 μW, at which net stimulated gain is vanished. At transparent pump power, a cold quality factor of approximately 1,500 is estimated. We did not find lasing mode in other samples with different twisting angle (see supporting information), possibly due to relatively small quality factors.

## 3. Conclusion

In summary, we demonstrate a laser based on photonic AQTIs. We design the PhCs with non-trivial quadrupole topological phases by twisting the holes within each unit cell.

The corner state, which is topologically protected, emerging at the interface between two slab PhCs with opposite twisting angle. The quality factor and mode volume of corner states can be strongly modified by twisting angle. Experimentally, we observed strongly localized corner states with high quality factor around 1,500. With increasing pump power, the corner state shows a transition into lasing oscillation at a threshold of approximately 23 μW. A high $β$ factor of 0.15 is observed, indicating the small mode volume of corner mode. Our results show the first observation of a corner state lasing in AQTI phase in semiconductor integrated photonics platform, and pave a novel way towards topological laser and topological integrated photonic platforms.

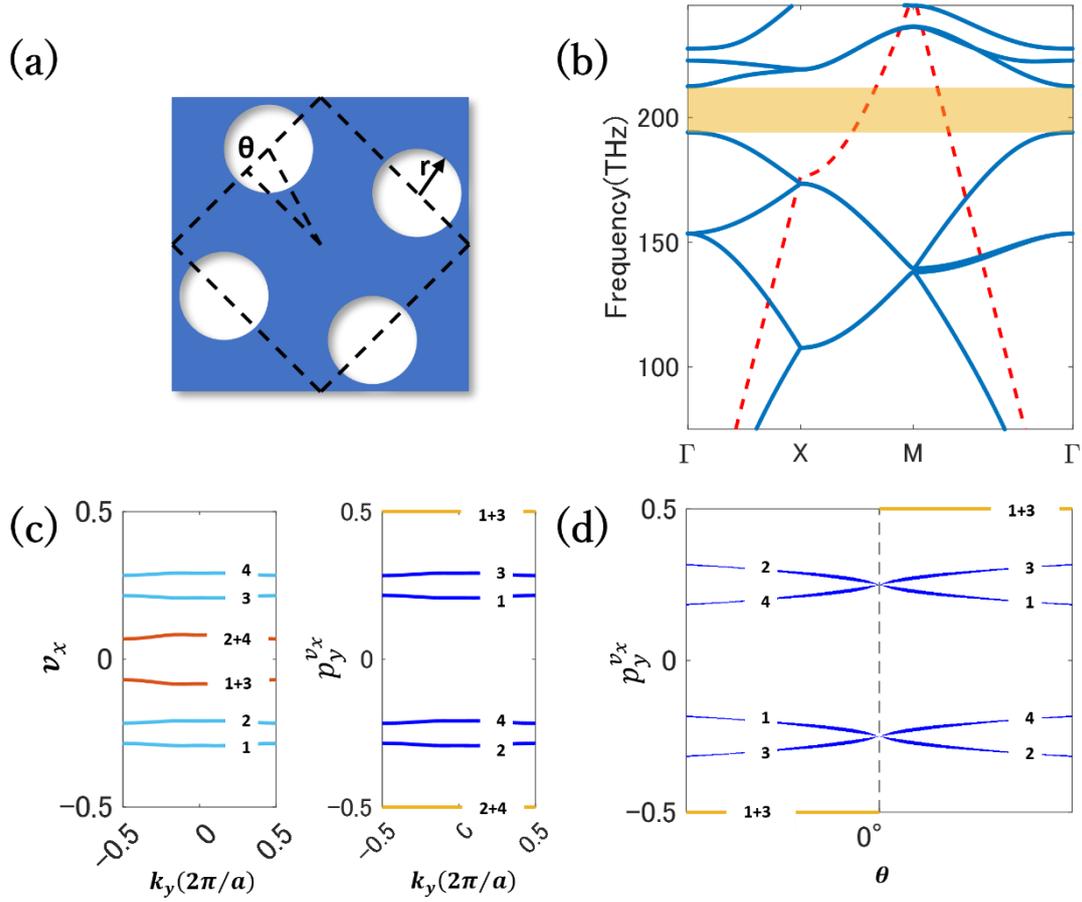

Figure 1. a) Schematic of photonic anomalous quadrupole topological insulators. The twisting angle (where a positive value indicates a clockwise direction) is represented by the symbol θ. b) Photonic TE band diagram for $\theta = 6°$ calculated by 3D-FEM. The red dashed line indicating the light line and yellow region corresponds to a topological non-trivial band gap. c) Wannier bands and nested Wannier bands for $\theta = 6°$. Two Wannier sectors I and II, labeled as 1+3 and 2+4, have gapped Wannier bands and quantized edge polarizations, indicating the non-trivial quadrupole topological phase. d) Evolutions of nested Wannier bands with the twisting angle θ. Phase transition happens at $\theta = 0°$.

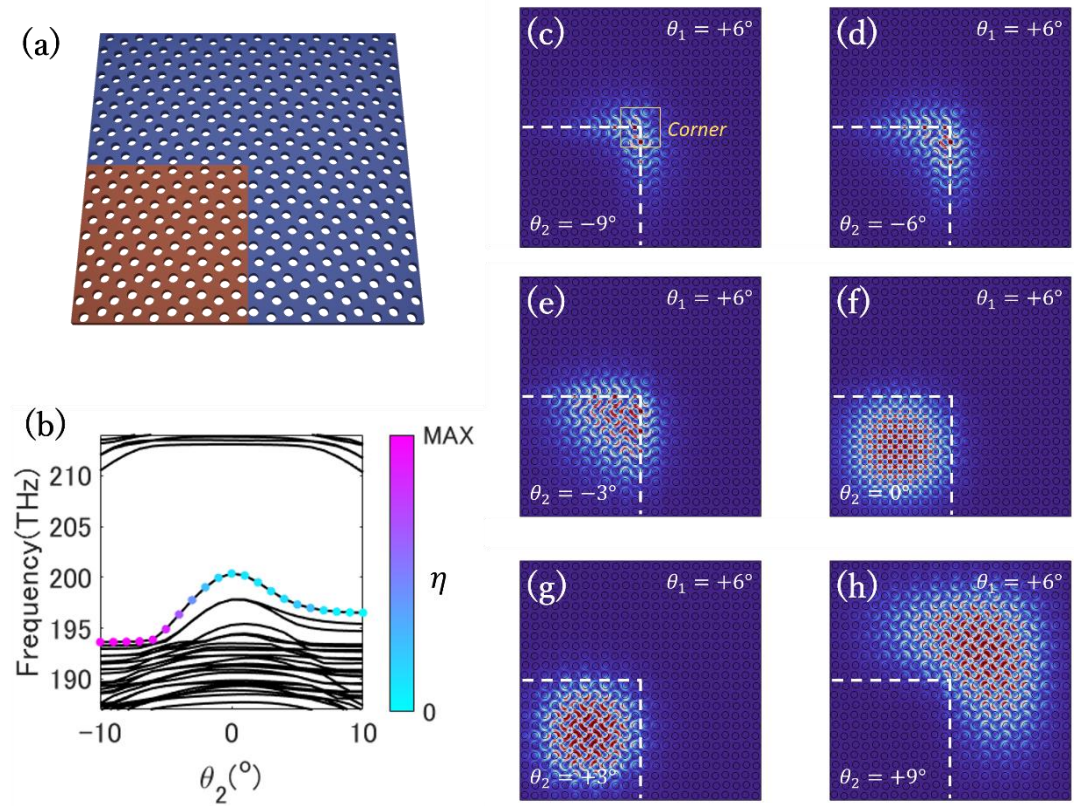

Figure 2 a) Schematic illustration of the investigated cavity. Blue and red regions represents PhCs with different twisting angles. b) Black lines shows the frequency evolution of eigenstates. The localization of states are shown by the color of dots. c-h) Evolution of field profile for corner state with different twisting angle. c-e) Corner state exists when the sign of twisting angles are different. f-h) Corner state disappears when one of the twisting angles is zero or two twisting angles have the same sign.

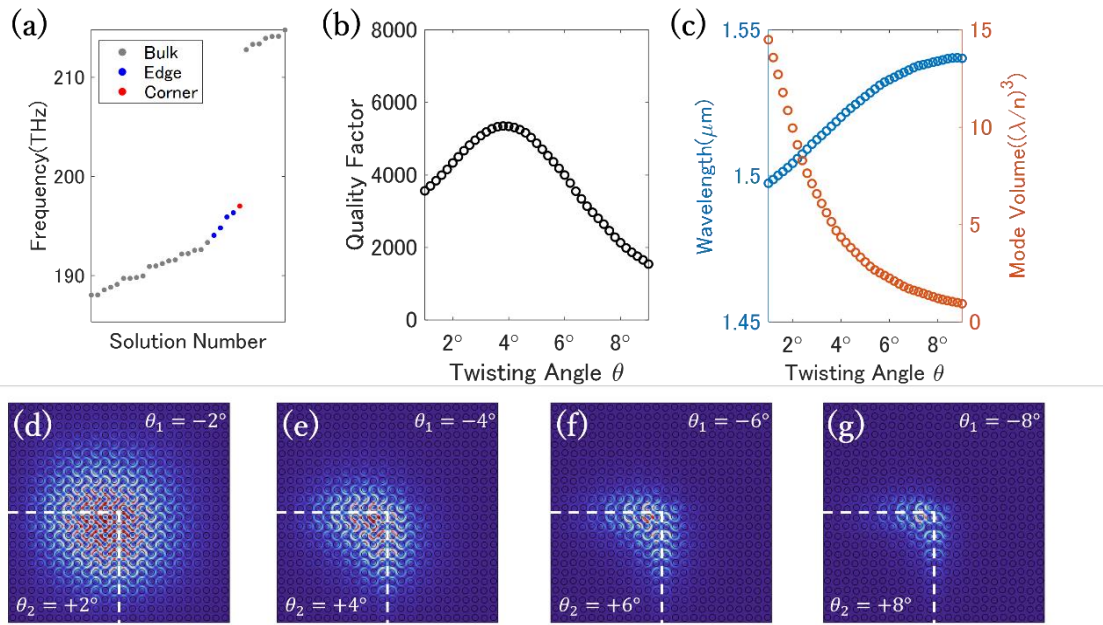

Figure 3. a) Eigenfrequency of the investigated cavity with $\theta = 6°$. A single corner state is found within the bandgap. b-c) Evolution of b) quality factor, c) frequency and mode volume of corner state as a function of twisting angle $\theta$. d-g) Filed profile of corner states with different twisting angles. Corner state becomes more localized as twisting angle increases.

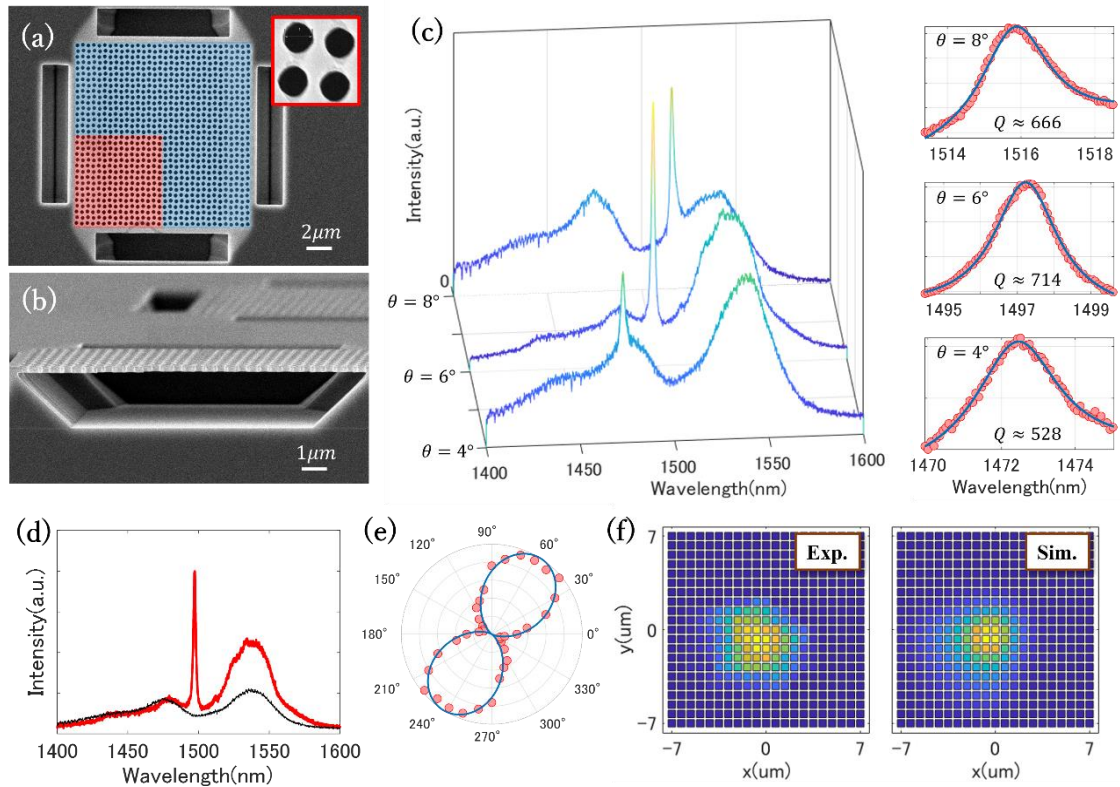

Figure 4. Scanning electron microscope image of a) the top view and b) the side view of a fabricated sample with $\theta = 6º$. The red and blue regions signify two PhCs with opposing twisting angles. The red box shows an enlarged unit cell c) Left, photoluminescence spectra for samples with different twisting angles. Right, close up spectra of cavity modes for different twisting angles. The blue curves are fitting curves. d-f) Optical characterization for sample with $\theta = 6º$. d) Photoluminescence spectra for structures composed of two topologically distinct PhCs (red line) and a single PhCs (black line). The resonant peak disappears in black lines. e) Polarization of the cavity mode. The blue curve is the fitting to the data points. The direction is defined to be same as (a). f) Experiment and simulation results for spatial extends of the cavity mode. The simulation result is the convolution of the field distribution and a gaussian beam.

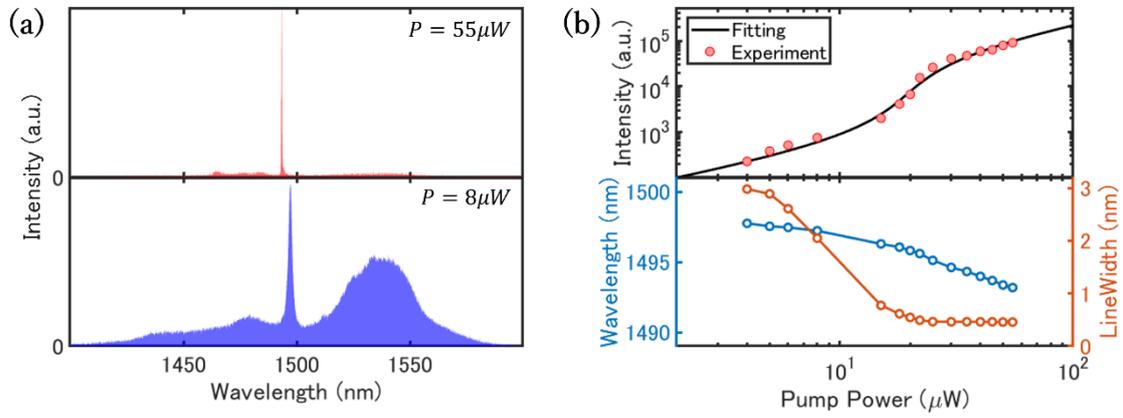

Figure 5. Lasing oscillation from the sample. a) Photoluminescence spectra for two different pump powers. Top, above the threshold. Bottom, below the threshold. b) Top, logarithmic plot of the LL curve. The mild S-shape indicating a high-β laser. Bottom, evolution of wavelength and linewidth as a function of pump power.

**Supporting Information**

Supporting Information is available from the Wiley Online Library or from the author.

**Acknowledgements**

The authors thank R. Miyazaki, Dr. S. Ishida, M. Nishioka, Dr S. Ji, and Dr. N. Ishida for their technical support and helpful discussions. This work was supported by JST CREST(JPMJCR19T1), KAKENHI (22H00298, 22H01994), Asahi Glass Foundation.